\documentstyle[]{laa}
\begin{document}
 
  \thesaurus{09     % A&A Section 6: Form. struct. and evolut. of stars
       (09.03.1; % 
	09.09.1; % 
	09.10.1; % 
	08.06.2; %
	08.13.2) % 
       }
  \title{Jets and high-velocity bullets in the Orion A outflows. Is the IRc2 outflow 
powered by a variable jet?}
 
%  \subtitle{}
 
  \author{A. Rodr\'{\i}guez$-$Franco$^{1,2}$, J. Mart\'{\i}n$-$Pintado$^2$, and
	  T.L. Wilson$^{3,4}$
%     \inst{1}
     }

  \offprints{A. Rodr\'{\i}guez$-$Franco to the IGN address}
 
  \institute{$^1$Departamento de Matem\'atica Aplicada$\,$II, Secci\'on 
       departamental de Optica, Escuela Universitaria de Optica, 
       Universidad Complutense de Madrid. Av. Arcos de Jal\'on s/n. 
       E-28037 Madrid, Spain\\
       $^2$Observatorio Astron\'omico Nacional (IGN),
       Campus Universitario, Apdo. 1143, E-28800, 
       Alcal\'a de Henares, Spain\\
       $^3$Max Planck Institut f\"ur Radioastronomie, Postfach 2024, D-53010 Bonn,
       Germany\\
       $^4$Sub-mm Telescope Observatory, Steward Observatory, The University of Arizona,
       Tucson, Az, 85721
       } 

  \date{Received, 1998; accepted, 1999}
 
  \maketitle
 
  \begin{abstract}
%______________________________________ Do not leave a blank line here!
%
We present high sensitivity maps of the High Velocity (HV) CO emission 
toward the
molecular outflows around IRc2 and Orion--S in the Orion A molecular cloud.
The maps reveal the presence of HV bullets in both outflows with velocities between 
40-80$\,$km$\,$s$^{-1}$ 
from the ambient gas velocity. The blue and redshifted CO HV bullets 
associated with the IRc2 outflow 
are distributed in thin ($12''-20''$, $0.02-0.04\,$pc) elliptical
ring-like structures with a size of $\sim10''\times50''$ ($0.02\times0.1\,$pc). The CO 
emission
at the most extreme blue and redshifted
velocities (EHV) peaks $20''$ north of source I, just inside the rings of the 
HV bullets. 
  
The low 
velocity H$_2$O masers and the H$_2^*$ bullets around IRc2 are located at the 
inner edges of the ring of CO HV bullets and surrounding the EHV CO 
emission. Furthermore,  the high velocity H$_2$O
masers are very well correlated with the EHV CO emission. This morphology is consistent 
with a model of a jet driven molecular outflow 
oriented close to the line of sight.  

In the Orion--S outflow, 
the morphology of the CO HV bullets shows  
a bipolar structure in the southeast$\leftrightarrow$northwest direction, and
the H$_2$O masers are found only at low velocities in the region between the
exciting source and the CO HV bullets. 

The morphology of the CO HV bullets, the radial velocities and the spatial 
distribution of the H$_2$O masers in both outflows, as well as the H$_2^*$ features 
around IRc2, are consistent with a model in which these outflows are driven 
by a jet variable in direction.
In this scenario, the large traverse velocity measured for the H$_2$O masers in the IRc2 
outflow, 
$\sim18\,$km$\,$s$^{-1}$, supports the evolutionary connection between the jet 
and the shell-like outflows. 
%_____________________________________ Do not leave a blank line here!
    \keywords{ISM: clouds --
	ISM: jets and outflows --
	nebulae: Orion Nebula --
	Stars: formation --
	Stars: mass-loss --
	shock waves
	}
  \end{abstract}
 
%
%________________________________________________________________
 
\section{Introduction}

Molecular outflows associated with young stellar objects are mostly made of ambient 
material entrained by a primary wind from the central source. While young 
molecular outflows are highly collimated with HV jets (see e.g. Bachiller, 1996), 
more evolved outflows are poorly collimated with shell-like 
structures (see e.g. Snell et al., 1980; 
Fuente et al., 1998). Different kinds of models (wind-driven bubbles and steady state jets)  
have been developed  to explain the two types of outflows (see e.g. Cabrit, 1995) but none of them can 
account for all the observational properties. 
A jet variable in velocity and/or direction, 
would explain the momentum distribution (Chernin \& Masson, 1995), the multiple 
acceleration sites (see e.g. Bachiller, 1996), and the evidences of a wiggling
molecular outflow (Davis et al., 1997). Furthermore,
models of the interaction of a jet variable in time and 
direction predict that the jet breaks, given rise to independent HV
bullets located in a shell-like structure with 
a non-negligible transverse velocity component (see e.g. Raga \& Biro, 1993). 
Thus, observations of molecular outflows oriented along the line of sight and powered by a 
variable jet
should show a ring-like structure of HV jet-bullets, and would allow to test these kind
of models.

In this 
letter, we present 
high sensitivity CO observations of the molecular outflows in the Orion A molecular cloud
(IRc2, see e.g. Wilson et al.,1986; and Orion-S, Schmid$-$Burgk et al., 1990).
The morphology of the HV CO emission around the IRc2 outflow reveals the presence of
a bipolar structure $20''$ from I source surrounded by a HV,
ring-like structure of CO bullets that cannot be
explained by the weakly collimated bipolar outflow model proposed by Chernin \& 
Wright (1996), and suggest a jet driven molecular outflow oriented along the line of sight
(Johnston et al., 1992).  
The combination of our results with those of the H$_2$O masers, strongly supports 
the idea that these molecular outflows are driven by jets which change in 
direction with time.

%
%__________________________________________________________________
 
\section{Observations}

The observations of the $J=2\rightarrow1$ lines of 
CO where carried out with the IRAM 30-m telescope at Pico Veleta (Spain). 
The observation were made with a SIS 
receiver tuned to single side band (SSB) with an image rejection of 
$\sim 8\,$dB. The SSB noise temperatures of the receivers at the rest 
frequency was 300$\,$K. The half power beam width of the
telescope was $12''$. For spectrometers, we used a filter banks of 
$512\times1\,$MHz that provided a velocity resolution of 
1.3$\,$km$\,$s$^{-1}$. The observation procedure was position switching 
with the reference taken at a fixed position located 15$'$ away in right 
ascension. The mapping was carried out by combining 5 on-source spectra 
with one reference spectrum. The typical integration times were 20$\,$sec for 
the on-positions and 45$\,$sec for the reference spectra. Pointing was checked 
frequently on nearby continuum sources and Jupiter, and the pointing errors were 
$\leq4''$. The calibration of the data was made by observing the sky, a hot 
and a cold load. The line intensity scale has been converted to units of main 
beam brightness temperatures by using a main beam efficiency of 0.45. The RMS noise of 
the map was 0.6$\,$K.

\section{HV bullets in the IRc2 and Orion-S outflows}

\subsection{Spectral features}

  \begin{figure*}
   \vspace{8.5cm}
   \includegraphics{ori-lan-f1.ps}
   \caption{Left panels: a) sample CO $J=2\rightarrow1$ line 
	profiles taken towards 
	selected positions in the vicinity of the molecular outflow
	surrounding IRc2. The offsets shown in the upper right corner of each box 
	are relative to IRc2. The vertical arrows show 
	the location of HV ``bullets'' similar 
	to those observed in some bipolar outflows driven by low mass stars.
	Central panels: b) and c) integrated intensity of the CO $J=2\rightarrow1$ line
	emission
	between $-90$ and $-10\,$km$\,$s$^{-1}$, and between 
	$40$ and $90\,$km$\,$s$^{-1}$ for the blue and red HV bullets 
	respectively. The maps have
	been obtained by subtracting a Gaussian profile to the broad line wings.
	The first contours level is $2\,$K$\,$km$\,$s$^{-1}$, and the interval between
	levels is $7\,$K$\,$km$\,$s$^{-1}$.
	d) and e) integrated intensity maps of the CO $J=2\rightarrow1$ line over the
	most extreme velocities (``molecular jet''), from $-110$ to
	$-90\,$km$\,$s$^{-1}$ for the blue jet, and from 
	$95$ to $115\,$km$\,$s$^{-1}$ for the red jet. For these two panels,
	the first contour level
	is $2\,$K$\,$km$\,$s$^{-1}$ and the interval between
	levels is $1\,$K$\,$km$\,$s$^{-1}$.
	The circle in the lower left panel
	represents the size of the beam. For all the panels,
	the filled star represents the position of IRc2, and triangles and dots represent, 
	respectively, the
	positions of the high and low velocity H$_2$O maser taken from Gaume et al.
	(1998), and the filled squares the positions of some H$_2^*$ features (Stolovy
	et al., 1998).} 
     \label{fig:ring}
  \end{figure*}

The left panel of Fig. \ref{fig:ring} (panel a) shows a sample of spectra 
taken toward the IRc2 outflow. The profiles show the typical broad line wings 
($\pm 100\,$km$\,$s$^{-1}$) associated with this molecular outflow.
Superposed on these, because of the better sensitivity than previous published data,
we have detected  well defined spectral HV
features restricted to certain radial velocity ranges (see the vertical arrows 
in the spectra of Fig. \ref{fig:ring}a). 
Most of the HV features in the IRc2 outflow appear at radial velocities between 30 and 
$90\,$km$\,$s$^{-1}$. Similar HV features are also clearly identified in the spectra of 
Orion-S outflow (left panel of Fig. \ref{fig:ori-s}; Schmid-Burgk, private communication ).
The HV spectral features detected in both molecular outflows are reminiscent of 
the HV bullets found in the molecular outflows driven by low mass stars
(see Bachiller, 1996). Furthermore, the CO 
profiles towards the IRc2 outflow  are similar to  the 
recently discovered H$_2^*$ bullets (Stolovy et al., 1998). The CO HV features reported
here 
represents the first detection of HV bullets in molecular outflow powered 
by a massive star.

\subsection{Morphology}

The upper right panels (b and c) of Fig. \ref{fig:ring}, and the right panel 
of Fig. \ref{fig:ori-s} show, respectively, the spatial distribution of the 
integrated intensity of the 
blue and redshifted CO HV bullets in the IRc2 and Orion-S outflows.
The spatial distribution of the HV bullets have been obtained by subtracting 
the smooth broad velocity  component by fitting a Gaussian profile.
It is remarkable that the blue and redshifted CO HV bullets around IRc2 
(Fig. \ref{fig:ring}b, and c) are distributed in an elliptical ring-like structure 
with a 
size of $\sim 10''\times50''$ (0.02$\times0.1\,$pc at the distance of 0.5$\,$Kpc) 
with IRc2 located in the southeast edge of the
HV bullets rings.
The ring morphology of the CO HV bullets in the IRc2 outflow shows only 
minor changes with the radial velocity, 
indicating a nearly uniform distribution over the whole velocity 
range.
Although the typical 
thickness of the rings is $12''-20''$ ($0.02-0.04\,$pc), some positions
are unresolved (thicknesses $\leq6''$; 0.01$\,$pc). This suggests that 
the blue and redshifted 
CO HV bullets are generated in a thin layer of HV gas distributed in a ring-like 
structure. It is interesting to note that the HV bullet rings are broken in the northwest
edge just at the bottom of the H$^*_2$ fingers (Allen \& Burton, 1993). This
indicates that the H$^*_2$ fingers might have been produced when the hot gas within the 
shocked
region breaks into the more diffuse medium and rapidly expands (McCaughrean \& Mac Low,
1997).

In panels d and e of Fig. \ref{fig:ring} we also show the spatial distribution 
of the CO emission for the most EHV components ($\geq\mid90\mid\,$km$\,$s$^{-1}$). 
The bulk 
of the blueshifted and 
redshifted EHV gas is located $20''$ north of source I. None of the current jet and wind 
models can account of all the observational 
properties of the IRc2 outflow. The morphology of the CO emission at moderate 
velocities favors a  biconical outflow structure that has a wide ($130^{\rm o}$) 
opening angle (Chernin \& Wright, 1996) powered by source I (Menten \& Reid, 1995). 
The morphology of the SiO maser spots 
near source I is consistent with a wide angle biconical outflow, but this simple model 
cannot 
account for the H$_2$O maser emission (Greenhill et al., 1998; Doeleman et al., 1999). 
The morphology of the HV CO bullet ring-like structures roughly 
trace the edges of the proposed biconical structures. However, the bipolar distribution
of the EHV gas 
emission $20''$ north of source I, surrounded by the CO HV bullet rings
is inconsistent with a wide angle biconical outflow model. 
In the next section, we analyze the alternative model of 
a jet driven molecular outflow directed along the line of sight 
(Johnston et al., 1992).

Fig. \ref{fig:ori-s}b shows the spatial distribution of the HV bullets for the 
Orion--S outflow. The HV bullets are small 
condensations (size $\sim30''$; 0.07$\,$pc) and
show the typical bipolar distribution with the blue and redshifted 
features spatially separated from the powering source. The HV bullets basically outlines 
the full extent of this outflow. The outflow containing the CO bullets 
reported in this letter is perpendicular to the low velocity redshifted 
outflow 
found by Schmid-Burgk et al. (1990), and consistent with the spatial distribution
of the SiO outflow
found by Ziurys et al. (1990). 
The possible driving source, as defined by the center 
of symmetry from the kinematics of the HV gas, must lie at a position 
$\sim18''$ north from FIR$\,4$ where no prominent continuum source has been 
detected so far (Mundy et al., 1986; Wilson et al., 1986).
In this outflow, the blue HV bullet shows different 
spatial distribution and velocity than the redshifted one.
While the red CO bullet has a moderate radial velocities of $\sim60\,$km$\,$s$^{-1}$
and is
close to the exciting source, the blue CO bullet has very high velocity
($\sim100\,$km$\,$s$^{-1}$) and is located further away from the exciting source. The
different distribution might be due to the fact that the blue bullet is less massive 
than the red one, and it 
has been already accelerated up to the terminal velocity. 
In fact, the Orion--S outflow is rather young with a dynamical age of only $10^3\,$years.

\section{Discussion}

\subsection{Jet driven molecular outflow in Orion A}

With the present data we conclude that the CO HV bullet rings around IRc2 
represent thin layers of HV condensations which have been shocked and 
accelerated by a fast jet oriented along the line of sight.
The orientation of the flow along the line of sight is also suggested by the kinematics
of the SiO masers around source I (Doeleman et al., 1999). In addition
to the morphological arguments, the 
location of different shock tracers in this region like the H$_2$O masers, their
kinematics, 
and the H$_2^*$ features can be accounted by this model.
Fig. \ref{fig:ring}b, c, d and e, show the location of the low (filled circles), the
high velocity (open triangles) H$_2$O masers, and the H$_2^*$ 
bullets (filled squares) adapted from Stolovy et 
al. (1998). The shock tracers, the low velocity H$_2$O masers and the 
H$_2^*$ bullets, are located in the inner border of the
CO HV bullets ring and, therefore, are surrounding the EHV jet. 
The H$_2^*$ bullets and the H$_2$O masers are displaced from the CO HV bullets 
by $\sim10''$ 
($2\times10^{-2}\,$pc). On the other hand, the high velocity H$_2$O masers are 
well correlated with the EHV jet indicating that they, indeed, arise from the 
interaction of the jet with gas which was already accelerated close to 
the terminal velocity of the outflow.

\begin{figure}
   \vspace{4.3cm}
   \includegraphics{ori-lan-f2.ps}
   \caption{Left panels: a) sample of CO $J=2\rightarrow1$ spectra taken towards 
	selected positions of the molecular outflow Orion--S. The offsets shown 
	in the upper right corner of each box 
	are relative to the position of FIR$\,$4. The vertical arrows show 
	the location of HV ``bullets''. Right panel: b) integrated intensity maps 
	of the CO $J=2\rightarrow1$ line in the Orion--S outflow. 
	The offsets are relative to the position of FIR$\,$4.
	The intervals of
	velocity integration are: from $-150$ to $-5\,$km$\,$s$^{-1}$
	for the blue wing, and from 25 to 100$\,$km$\,$s$^{-1}$ for the red wing.
	For both wings the first contour level
	is $9\,$K$\,$km$\,$s$^{-1}$ and the interval between
	levels is $2.5\,$K$\,$km$\,$s$^{-1}$. The circle in the lower left panel
	represents the size of the beam. The filled triangles and dots represent, 
	respectively, the
	positions of the blue and red H$_2$O maser taken from Gaume et al. (1998).
	The filled square shows the position of the possible exciting source.} 
     \label{fig:ori-s}
  \end{figure}

The observed morphologies of the CO bullet ring, the shock tracers, and 
the EHV gas can be explained  by a fast jet moving along the line of sight and
interacting with the surrounding molecular gas. In this 
scenario, the fast jet with material moving at velocities
$\geq100\,$km$\,$s$^{-1}$ will interact with the surrounding ambient 
gas generating strong shocks 
that will compress, heat the gas, and even
photodissociate molecules in its surroundings and in the head of the jet.  
As one moves away from the working surfaces of the jet, the material will cool down.
First, H$_2$ molecules will be formed producing the strong H$_2^*$ features 
in the densest hot clumps, and also the H$_2$O maser emission. 
As the accelerated post-shock material moves further away from the interface region 
it will cool further forming CO molecules (see Hollenbach \& McKee, 1989) and 
given rise to the CO HV bullets.

The stratification of the different shock tracers indicates that
the shocked 
gas is not only accelerated along the line of sight, but also
perpendicularly to the jet axis. This is consistent with the 
measured proper motion of the low velocity H$_2$O 
masers which indicates that these are expanding at a velocity of 
$18\,$km$\,$s$^{-1}$ (Genzel et al., 1981). 
This allows us to measure for the first time the transverse velocity in a jet driven
molecular outflow which is
$\sim20\%$ of the jet velocity.
For a transverse velocity of $18\,$km$\,$s$^{-1}$, the separation between the 
CO bullet ring and the H$_2^*$ indicates that the CO bullets ring has gone
through the shock $\sim10^3\,$years ago, which is at least one order of magnitude 
larger than the 
typical time required to cool down the material and to produce CO molecules
efficiently (Hollenbach \& McKee, 1989). This indicates that
the H$_2^*$ and the H$_2$O emissions trace very recent shocks 
produced
less than 100$\,$years ago, and the CO HV bullets must have been produced by shocks more
than $10^3\,$years ago.

In contrast to the IRc2 outflow, the Orion--S outflow seems to be 
oriented nearly 
perpendicular to the line of sight.
The combination of these two outflows powered by massive stars with strong H$_2$O maser
emission, but oriented with very different
angles to the line of sight allows a three dimensional study of 
the interaction of the HV jets with the ambient cloud.

We now study the history of the young Orion--S outflow using the picture of the time 
dependent shock tracers obtained from the IRc2 outflow. In the case of the Orion--S 
outflow, all the H$_2$O masers are associated with the 
HV gas seen in CO (lower panel of Fig. \ref{fig:ori-s}b). The redshifted H$_2$O
masers are on the western part of the  
red bullet facing the exciting source where a jet, nearly 
perpendicular to the line of sight, impinges. The interaction of the jet 
with the CO HV bullets is not only supported by the morphology, but also by 
the H$_2$O masers which have radial velocities similar to that of the red CO 
bullet. This
indicates that the most recent shocks are indeed produced in the HV bullet. 
The blueshifted bullet seems to be in a different stage of evolution. 
As previously mentioned, the blue bullet is further from the 
exciting source than the red one, and the blueshifted masers are only found close to 
the exciting source at radial velocities close to the ambient velocities. This 
indicates that the 
most recent interaction of the jet is not occurring with the HV bullet material, but in 
ambient material which has not been yet affected by the jet, suggesting that 
the jet might have slightly changed the direction at which it is ejected.
Therefore this can be understood in a 
model in which the molecular outflows are powered by a jet whose 
direction is changing with time (see e.g. Raga \& Biro, 1993).

\subsection{The Origin of shell-like molecular outflows}

The large amount of the molecular gas mass observed in outflows suggests that these are
mostly made by ambient material entrained by a ``primary jet'' (Bachiller, 1996).
Entrainment can be divided in two categories: prompt entrainment at the jet
heads (bowshock), and steady-state entrainment along the side of the jet due to
Kelvin-Helmholtz (KH) instabilities. Most of the studies of the lines profiles seem
to indicate that the prompt entrainment at the jet heads is the main mechanism for
molecular entrainment (Chernin et al., 1994). However, to explain the data, 
Chernin et al. (1994) proposed 
a jet/bowshock model in which the 
jet is variable in velocity and/or direction. In the proposed scenario of a jet 
along the line of sight powering the IRc2 outflow,  changes in the direction of 
the jet would explain the ring-like structure of the  
CO bullets (Raga \& Biro, 1993), the spatial distribution of all the
H$_2$O masers, as well as the large number of the H$_2$O masers at relatively 
low radial velocities. One important aspect of our interpretation
is that the entrained 
material is moving perpendicular to the jet with velocities of up to $20\%$ of 
the jet velocity.  This indicates that in a time scale of $10^5\,$years, the IRc2 
molecular outflow will form a cavity around the driving source with typical 
size of $\sim3\,$pc. This is similar to those found in molecular  
outflows  powered by intermediate mass stars (NGC$\,$7023; Fuente et al., 1998) 
and by low mass stars (L$\,$1551-IRS5; Snell et al., 1980).

In summary, the morphology and the radial velocities of the CO HV bullets,
the H$_2$O masers and the H$_2^*$ bullets in the Orion A outflows can be 
explained by the interaction of jet-driven molecular outflows powered by 
variable jets in direction with the ambient gas.

  \acknowledgements
%________________________________________ Do not leave a blank line here!
   We thank Dr. A. Fuente for  critical reading of the manuscript, and 
   the referee, Dr. D.S. Shepherd, for her suggestions.
   Part of this work was supported by the Spanish DGES under grant 
   number PB96-0104.
 
%
%_____________________________________________________________________


\begin{thebibliography}{}

  \bibitem{} Allen , D.A., Burton, M.G.: 1993, Nature, 363, 54. 

  \bibitem{} Bachiller, R.: 1996, ARA\&A 34, 111.

  \bibitem{} Cabrit, S.: 1995, A\&SS 233, 81.

  \bibitem{} Chernin, L.M., Masson, C.R., Gouveia dal Pino, E.M., Benz W.: 1994, 
	     ApJ 426, 204.

  \bibitem{} Chernin, L.M., Masson, C.R.: 1995, ApJ 455, 182.

  \bibitem{} Chernin, L.M., Wright, M.C.H.: 1996, ApJ 467, 676.

  \bibitem{} Davis, C.J., Eisl\"offel, J., Ray, T.P., Jenness, T.: 1997, 
	     A\&A 324, 1013.

  \bibitem{} Doeleman, S.S., Londsdale, C.J., Pelkey, S.: 1999, ApJL 510, L55.

  \bibitem{} Fuente, A., Mart\'{\i}n--Pintado, J., Rodr\'{\i}guez--Franco, A.,
	     Moriarty-Schieven, G.D.: 1998, A\&A  339, 575.

  \bibitem{} Gaume, R.A., Wilson, T.L., Vrba, F.J., Johnston, K.J., 
	     Schmid-Burgk, J.: 1998, ApJ 493, 940.
	     
  \bibitem{} Genzel , R., Reid, M.J., Moran, J.M., Downes, D.: 1981, ApJ 244, 884.	     

  \bibitem{} Greenhill, L.J., Gwinn, C.R., Schwartz, C., Moran, J.M., Diamond, P.J.: 
             1998, Nature 396, 650.

  \bibitem{} Hollenbach, D.J., McKee, C.F.: 1989, ApJ 342, 306.
	     
  \bibitem{} Johnston, K.J., Gaume, R., Stolovy, S., Wilson, T.L., Walmsley, 
	     C.M., Menten, K.M.: 1992, ApJ 385, 232.

  \bibitem{} Menten, K.M., Reid, M.J.: 1995, ApJLet 445, L157.

  \bibitem{} McCaughrean, M.J., Mac Low, M.M.: 1997, AJ 113, 391.

  \bibitem{} Mundy, L.G., Scoville, N.Z., B\aa \aa th, L.B., Masson, C.R., Woody, D.P.:
	     1986, ApJLet 304, L51.

  \bibitem{} Raga, A.C., Biro, S.: 1993, MNRAS 264, 758. 

  \bibitem{} Rodr\'{\i}guez--Franco, A., Mart\'{\i}n--Pintado, J., G\'omez--Gonz\'alez,
	     J., Planesas, P.: 1992, A\&A 264, 592.
  
  \bibitem{} Rodr\'{\i}guez--Franco, A., Mart\'{\i}n--Pintado, J., Fuente, A.: 1998,
	     A\&A 329, 1097.
	     
  \bibitem{} Schmid$-$Burgk, J., G\"usten, R., Mauersberger, R., Schulz, A., 
	     Wilson, T.L.: 1990, ApJLet 362, L25.

  \bibitem{} Snell, R.L., Loren, R.B., Plambeck, R.L.: 1980, ApJ 239, L17. 

  \bibitem{} Stolovy, S.R., Burton, M.G., et al.: 1998, ApJ 492, L151.
	    
  \bibitem{} Wilson, T.L., Serabyn, E., Henkel, C., Walmsley, C.M.: 1986, A\&A 158, L1.
  
  \bibitem{} Ziurys, L.M., Wilson, T.L., Mauersberger, R.: 1990, ApJLet 356, 
	     L25.
\end{thebibliography}
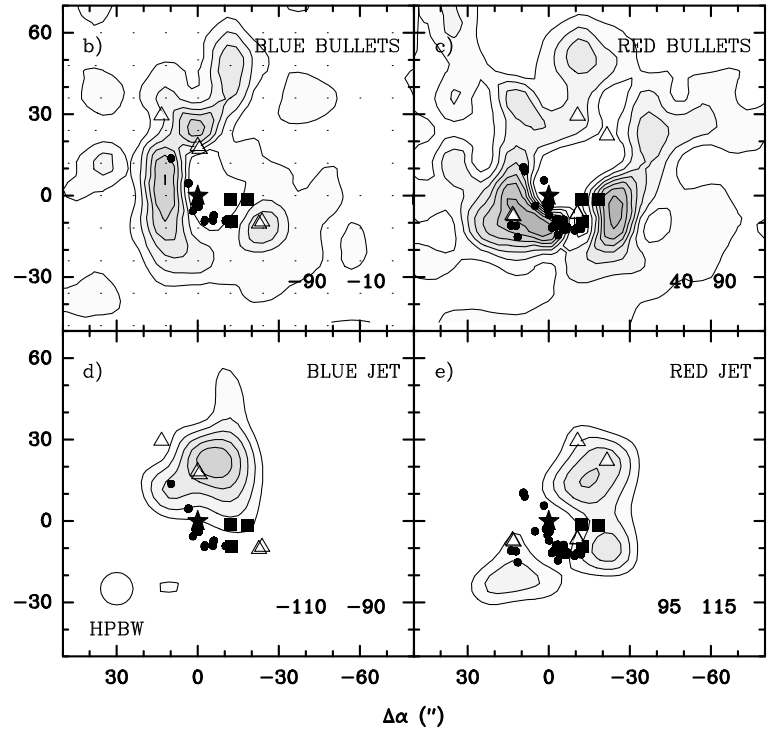
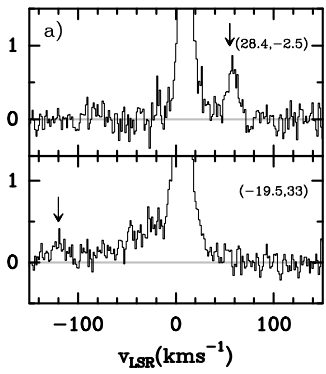
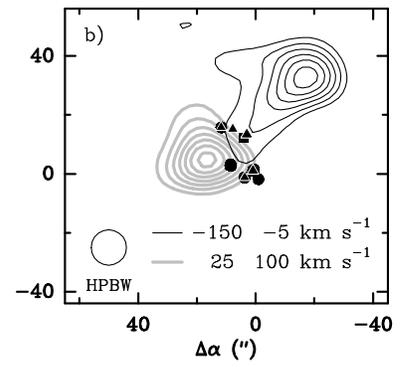
\end{document}